# Inverse Design of Inorganic Electrides

Yunwei Zhang[1], Hui Wang[1], Yanchao Wang [1], Lijun Zhang[1], and Yanming Ma[1,*]

[1]*State Key Lab of Superhard Materials, Jilin University, Changchun 130012, China*

*Address correspondence to: mym@calypso.cn

Electrides are ionic solids that consist of cationic frameworks and anionic electrons trapped in the voids of lattices. Organic electrides exist in a large abundance, but the thermal instability at room temperature and sensitivity to moisture are bottlenecks that limit their practical uses. Known inorganic electrides are rare but appear to have high thermal and chemical stability and exhibit promising applications as electron-emitting materials, superior catalysts and strong reducing agents. Here, we report a developed inverse-design method that can be used to search for a large variety of inorganic electrides. Our method utilizes the intrinsic property of interstitial electron localization of electrides as the global variable function being incorporated into the swarm-intelligence based structure searches. Through screening 99 binary ionic compounds, we have designed 89 new inorganic electrides that are classified into three-, two-, and zero-dimensional species according to the way that the interstitial electrons are localized and the conductive properties of the systems. Our work reveals the rich abundance of inorganic electrides by extending them into more general forms and provides new structure types for electrides that are not thought of as before.

Electrides are ionic solids with excess electrons trapped at the interstitial areas of lattices, which serve as anions[1-2]. The strongly or loosely localized electrons make electrides relate to salts with F-centers or plasmas, but markedly differ from metals containing delocalized electrons[3]. So far, most of the synthesized electrides are organic species composed of alkali metals ions complexed with crown ethers for the cations. The solvent electrons released by alkali metals are trapped in potential wells formed between these complexed cations[3,4,5,6]. Though they are reported as low-temperature electron emitters and strong reducing materials[4,5,7], the thermal instability above -40℃ and sensitivity to air and water restrict their practical application.

The first inorganic electride C12A7:$e^-$ was synthesized in 2003 by removing oxygen ions from the center of the clathrate Ca-Al-O cages of mayenite ($12CaO_7 \cdot Al_2O_3$)[8]. It shows superior properties that are thermally stable at room temperature and chemically stable in resisting to water and air. Excellent field-emission property is evident with even lower work function than carbon nanotubes[9]. An order of magnitude higher catalytic property than standard ones (such as iron-based catalyst $Fe\text{-}Al_2O_3\text{-}K_2O$[10]) is manifested when loading on Ru element in the ammonia synthesis[11].

Dicalcium nitride ($Ca_2N$) was recently proved to be a two-dimensional (2D) electride where anionic electrons are weakly localized in the 2D interspaces between two positively charged ionic $[Ca_2N]^+$ layers[12]. $Ca_2N$ was then used as an efficient electron donating agent in the transfer hydrogenation of alkynes and alkenes[13]. Monolayer $Ca_2N$ encapsulated by graphene shows excellent predicted electron transport properties if compared to typical 2D electronic systems (e.g., GaAs-AlGaAs heterojunction, $LaAlO_3$-$SrTiO_3$ interface and graphene)[14].

As an important category of inorganic electrides, high-pressure electrides were predicted in various simple elemental metals[15,16,17,18,19]. These elemental electrides are unique and differ from electron-rich ionic electrides seen at ambient pressure[20]. A drawback is that these high-pressure electrides are not quenchable to ambient conditions and hardly to be of any practical use.

Recently, a substitutional screening method[21] has been applied to search for 2D electrides. The rhombohedral anti-$CdCl_2$-type structure of $Ca_2N$ was used as a prototype electride structure where Ca and N were replaced with alternative metal elements (e.g., alkali and transition metals) and non-metal elements (e.g., halogen elements, O and C), respectively. On the one hand, the method is successful, especially when the potential electrides do crystalize in the same structure. There is the case on the prediction of $Y_2C$ electride that led to the actual experimental observation[22]. On the other hand, this method fails since it confines electrides in a certain structure type, which, however, limits our understanding of the diversity of electride structures. The predicted electrides strongly rely on the known structure types, while a finding of electrides with an unknown structure is impossible.

As described above, the inorganic electrides at ambient conditions are rarely known. This inevitably limits the wide range of applications of electrides. A massive computation-assisted design of inorganic electrides is highly desirable. Here we developed an efficient inverse-design method that relies on the intrinsic physical property of interstitial electron localization other than any prior known electride structures to search for a variety of new inorganic electrides. Our method has been incorporated into our in-house developed CALYPSO structure prediction code[23,24,25] enabling automatic structure searches through intelligent exploration of various unknown structure types. Our method is reliable to identify the known electrides of $Ca_2N$[12] and high pressure transparent Na[15] with the only given information of chemical compositions. Extensive inverse-design simulations on 99-targeted binary ionic systems were then performed, and 89 new electrides were successfully designed, some of which have already been synthesized, though not yet being pointed out as electrides, while others are new compounds awaiting experimental synthesis. We unraveled rich abundance of inorganic electrides by extending them into more general forms. A large variety of new prototype structures for electrides that are not thought of as before were reported.

## Results and Discussion

**Inverse-Design Methodology.** Our inverse-design method is on top of our developed swarm-intelligence based CALYPSO structure searching method that is able to intelligently explore structures with the only given information on chemical composition for a compound without relying on any prior known structural information[24]. In the current implementation, the degree of interstitial electron localization other than the total energy was introduced and adopted as the global variable function. This is in good accordance with the fact that strong electron localization in the voids of lattice is the intrinsic physical property of electrides. Our structural design follows the principle of inverse design where the structure searches are forced into a pursuit of a structure having a desirable functionality for a given chemical system. Our previous inverse-design exercise has been applied into the searching of superhard materials[26].

As a powerful technique to identify core, binding, and lone-pair regions in chemical systems, electron localization function (ELF)[27,28,29] provides a semi-quantitative index for the measure of interstitial electron localization. We define the degree of interstitial electron localization as the indicator to characterize the interstitially localized electrons. The indicator adopts the formula of $\omega = V_{inter}/V_{cry}$, where $V_{inter}$ and $V_{cry}$ are the volumes of the interstitial regions where electrons localize well and the simulation cell, respectively. $V_{inter}$ is critically determined from the reliable ELF calculation. A detailed description on how to obtain a precise interstitial $V_{inter}$ can be found in Supplementary Table S1 and Fig. S1. A larger ω gives a higher probability of the formation of electrides in a crystal. Though choices of ELF values for characterizing electron localization are dependent of specific systems, in our test, we found that ELF values larger than 0.75 can give a good description for electron localization for most of systems.

Our inverse-design approach is outlined in Fig.1a. In the first generation, all structures are generated randomly with the symmetry constraints. Structures are then geometrically optimized for seeking for their local minima in the potential surface. ω values of optimized structures are evaluated to rank good electride structures. A certain number of structures (here we chose 60% of the population size) with high-ω values are evolved into next generation by swarm-optimization operation via a smart learning of personal and global best electride structures[2330]. The rest 40% structures in each generation are randomly generated to enhance structural diversity during the structure evolution. Eventually, we are able to construct ω vs. energy map in an effort to trace the generated structures for acquiring desired electrides (e.g., Fig. 1b).

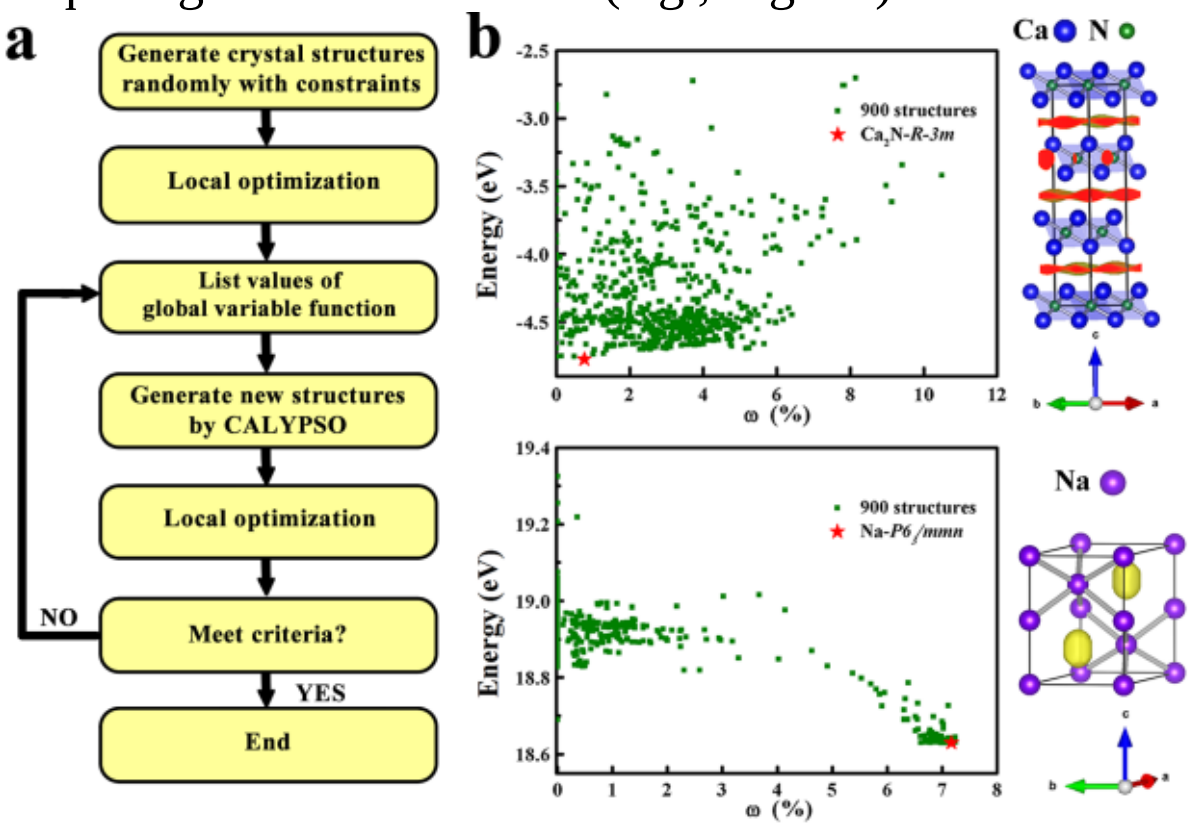

**Figure 1. Inverse-design scheme for electrides and its benchmarks on two known electrides of $Ca_2N$ and transparent Na at 320 GPa.** (a) Flow chart of CALYPSO module for searching for electrides by the inverse-design scheme. (b) ω vs. energy maps of $Ca_2N$ at ambient pressure (top left panel) and elemental sodium at 320 GPa (bottom left panel). Green squares represent structures produced by CALYPSO run within 30 generations. The experimental anti-$CdCl_2$ and hP4 electride structures for $Ca_2N$ and sodium are shown as red stars in left panels and depicted in top and bottom right panels, respectively.

**Benchmarks on Known Electrides.** We have benchmarked our inverse-design method onto two known electride systems: $Ca_2N$ (and its variants, such as $Sr_2N$ and $Y_2C$, shown in Supplementary Fig. S2) at ambient pressure and high-pressure electride of transparent sodium at 320 GPa. With the only input information of chemical compositions of Ca:N = 2:1, our calculations readily reproduced the experimental electride anti-$CdCl_2$-type structure of $Ca_2N$ as seen in Fig. 1b (top right panel) via the calculated ω vs. energy map (top left panel of Fig. 1b). Our calculations also correctly reproduced the hP4 electride structure (bottom right panel of Fig. 1b) of sodium at 320 GPa as seen in the calculated ω vs. energy map of Fig 1b (bottom left panel). This certifies the validity of our method in application to design of high-pressure electrides. Note that the valence electrons of sodium at 320 GPa are completely localized in the voids of hP4 structure that leads to the formation of an insulating electride[15].

**Two Design Principles for Inorganic Electrides.** It is known that electrides, such as $[Ca_{24}Al_{28}O_{64}]^{4+}$ $(4e^-)$ and $[M_2N]^+$ $(e^-)$ (M = Ca, Sr and Ba), contain intrinsic excess of electrons. Existence of excess electrons in a chemical system should be regarded as a necessary condition for stabilizing an electride and therefore has been used as one of our design principles in searching for inorganic electrides. We further prove this design principle in a model system by removing one excess electron per formula unit of $Ca_2N$. Upon depletion of its excess electron $Ca_2N$ is not an electride any more without showing any electron localization between the positively charged $[Ca_2N]^+$ layers (shown in Fig. 2a). In below context, we define the electron-rich binary systems into a general form of $[A_xB_y]^{n+}$ $(ne)^{n-}$ (x, y and n are integers; A and B are cationic and anionic elements as electrons donor and acceptor, respectively) for searching for potential electrides. These electron-rich ionic systems are in contrast to those conventional ionic solids (e.g., $Ca_3N_2$, YN and NaCl, etc.) that follow the rule of electroneutrality where the sum of the formal charges of all constituent elements is equal to zero.

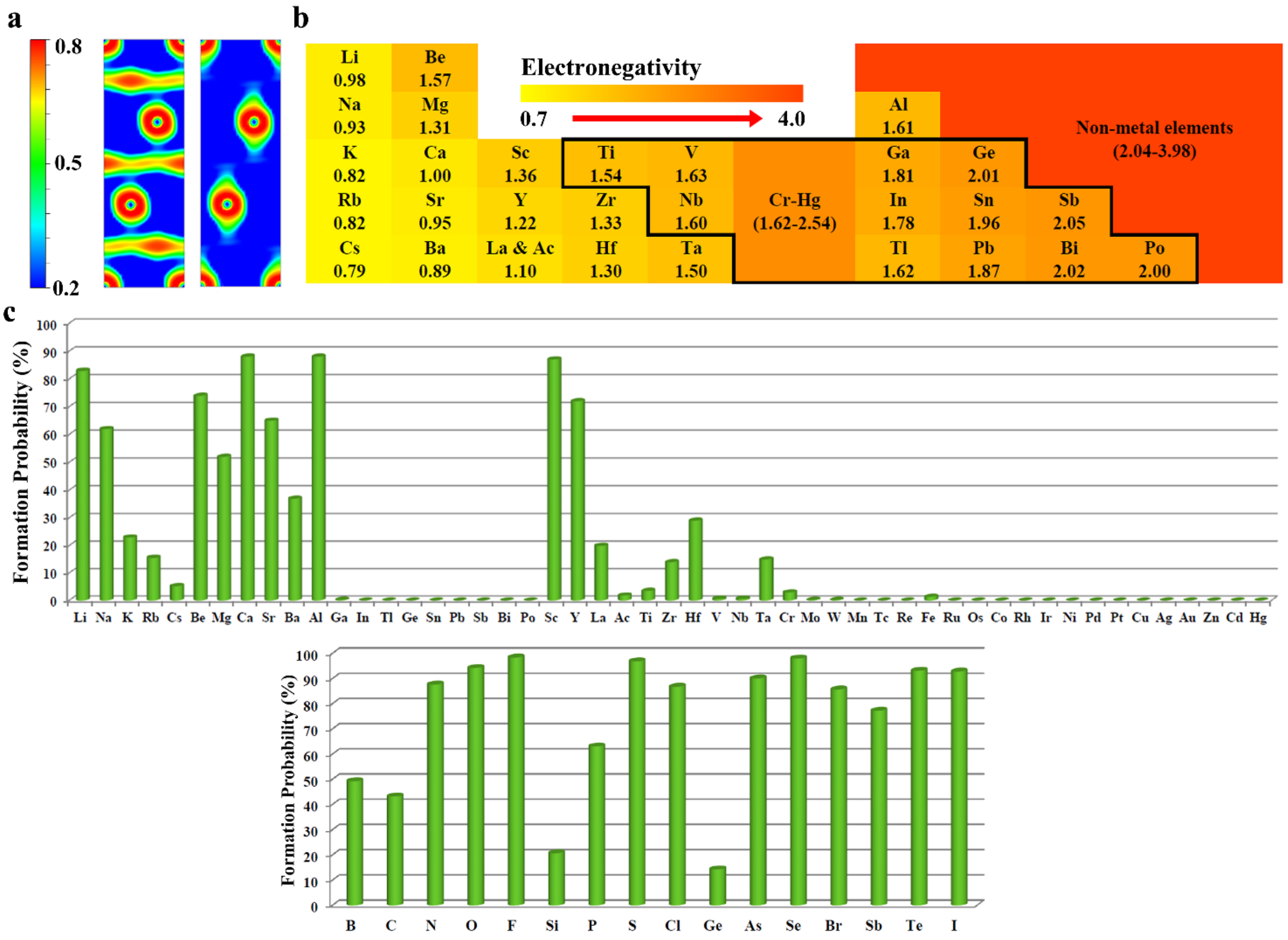

**Figure 2 | Tests for two design principles.** (a) The electron localization function of $Ca_2N$ (left panel) on the $(110)_R$ plane parallel to the hexagonal c-axis. Upon removal of one excess electron per $Ca_2N$, the system loses the feature of electron localization (right panel). (b) Electronegativity of elements by Pauling scale[31]. Metallic elements with higher electronegativity than Be are not good electron donors for binary electrides. (c) The formation probabilities (ratios of electride phases over all generated structures by CALYPSO code) of electrides in metal nitrides and $Ca_2X$ compounds are plotted in top and bottom panels, respectively. Metal and X elements are listed along x axis and formation probabilities of each compounds are shown as y axis.

Electronegativity of elements is another critical principle for the design of electrides since it is a direct measure of an element's ability to attract or donate electrons. We choose metal nitrides (La and Ac are representatives of lanthanides and actinides, respectively) and $Ca_2X$ (X = nonmetallic elements or metallic elements with high electronegativity comparable to nonmetals, such as Ge, Sb and Bi) as two model systems to illustrate how electronegativities of cationic and anionic elements can influence the formation of electrides in certain systems. Metal nitrides containing excess electrons, such as $[M_4N]^+ e^-$ for alkali metal nitrides and $[M_2N]^+ e^-$ for alkaline earth metal nitrides are examined (the results on other metal nitrides are listed in Supplementary Table S3). The formation probability of electrides, *i.e.*, the ratio of the generated electride structures over all structures produced by our inverse-design simulations, is plotted out in Fig. 2c. We found that only those metallic elements with low electronegativity are preferable to the formation of electrides. Specifically, nitrides formed by metallic elements in IA, IIA, and IIB groups and aluminum have higher formation probabilities of electrides, while the probabilities are much lower for IVB (e.g., Zr and Hf) elements. A few electrides in Ti and V nitrides appear, but they are energetically too unfavorable to allow the experimental synthesis. Other metallic elements having even higher electronegativity (indicated in the bold black frame in Fig. 2b) cannot form any electride phases.

For the $Ca_2X$ systems, we found most of testing objects have high formation probability of electrides beyond 50% (bottom panel in Fig. 2c), except for $Ca_2B$ and $Ca_2$-IVA compounds (e.g., $Ca_2C$, $Ca_2Si$ and $Ca_2Ge$). These latter four systems are not intrinsic electron-rich compounds through intuitive calculation of the sum of the formal charges of constituent elements. Most of the generated structures are therefore not electrides. Electride structures appear only when the anionic elements form pairs in the structure. Once paired, the resultant structure becomes an electron-rich system containing excess electrons satisfying the requirement for the formation of electride. Comparing the model calculation results of $Ca_2X$ system with those of metal nitrides, we found that the choice of cationic elements (other than the anionic elements) has a critical effect on the formation of electrides. Hence, it is desirable to confine our searching of electrides into those electron-rich systems composed of cationic elements with low electronegativity.

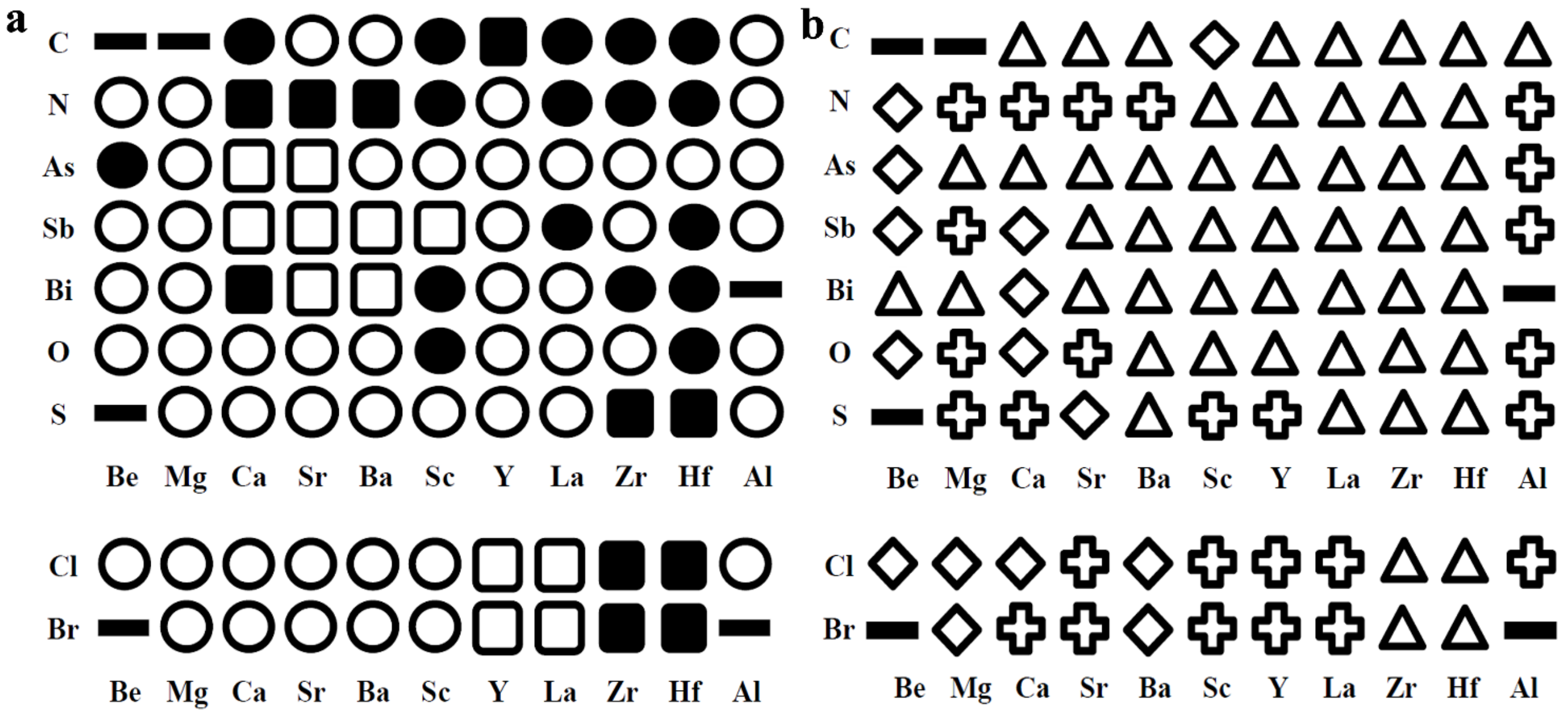

**Figure 3 | Inverse-design results of binary electrides.** (a) Stability Map of $A_2B$ (top) and AB (bottom) (A = electrons donor, B = electrons acceptor) electrides. Black lines indicate the absence of any reasonable electride phase in that system. Squares and circles denote the predicted electrides are existing compounds and yet to be synthesized, respectively. Solid symbols indicate the energetically stable phases (have negative formation energies) with respect to the existing compounds, while open symbols for metastable phases (have positive formation energies). (b) Type Map of $A_2B$ (top) and AB (bottom) electrides: triangles for 3D, diamonds for 0D, and crosses for 2D. Black lines indicate the same as in (a).

**New Electrides by Inverse-Design Simulations.** We conducted extensive inverse-design simulations through CALYPSO code on 99 electron-rich $A_2B$ and AB systems composed of 11 cationic A elements with low electronegativity playing the role of electron donors and 9 anionic B elements (7 non-metallic elements and 2 metallic elements with high electronegativity) as electron acceptors as shown in Fig. 3. Our calculations readily reproduced the correct anti-$CdCl_2$ structure shared by four known $Ca_2N$, $Sr_2N$, $Ba_2N$ and $Y_2C$ electrides[1,12,22]. We ruled out the formation possibility of electrides in nitrides and carbides dominating the hexagonal structure.

We here reported 89 newly designed electrides as depicted in the Stability Map (Fig. 3a). 3D ELF maps, structure information and formation energies of these electrides are shown in Supplementary Table S2. It is noted that for one particular system, many electride structures are generated in our simulation, however; only the electride having the lowest-energy structure has been presented in Fig. 3a. Among these 89 electrides, 19 are existing compounds (square symbols in Fig. 3a), but yet to be pointed out as electrides. The other 70 electrides (circle symbols in Fig. 3a) are hitherto unknown compounds awaiting experimental synthesis: 17 electrides are energetically stable (i.e., having negative formation energies) shown as solid circles in Fig. 3a, while the other 53 are metastable (i.e., having positive formation energies) shown as open circles. It is noteworthy that our target systems contain intrinsic excess electrons and therefore go against the electroneutrality for acquiring stable compounds. Chemically, this explains that some of our designed electrides are energetically metastable. In reality, these metastable electrides are experimentally synthesizable as seen in many examples on actual experimental syntheses (e.g., metastable compounds denoted by open squares in Fig.3a).

According to the way the excess electrons are localized and the conductive properties of systems, these electrides can be classified into three categories: 3D, 2D and 0D species as shown in Type Map (Fig. 3b), marked by different symbols. There are 52 3D, 22 2D, and 15 0D inorganic electrides, respectively. Below, we selected three compounds of $Ca_2C$, $Be_2N$ and LaCl as the illustrative examples for 3D, 0D and 2D electrides, respectively, to discuss their structural and electronic properties.

**Structures and Electronic Properties of Designed Electrides.**

**3D Electrides.** In 3D electrides (shown as triangles in Fig. 3b), excess electrons are partially localized in the cavity interstitials of the lattice. There is a subtle balance between localization and delocalization of excess electrons that contribute to the conductivity in three dimensions. $Ca_2C$ is a typical 3D electride that has a body-centered tetragonal *I4/mmm* structure (right panel in Fig. 4a) consisting of cationic $Ca^{2+}$ and anionic $C_2^{2-}$. This structure is a new prototype structure of electride that doesn't follow any known structures in the databases,[32,33] and it can be derived from the insulating calcium dicarbide $CaC_2$[34] (left panel in Fig. 4a) by replacing the dashed $C_2$ dimers with the interstitially localized electrons.

ELF results (right panel in Fig. 4b) showed that excess electrons are partially localized in anionic cavity sites that are linked each other by delocalized electrons. The calculated partial electron density for region near Fermi level (right panel in Fig. 4c) illustrates that these delocalized electrons dominate the conductivity of $Ca_2C$. The 3D conducting behavior of excess electrons can also be inferred from the band structure of $Ca_2C$ (left panel in Fig. 4c) at Fermi level that crosses over high symmetric directions along Z-A, A-M and X-Γ.

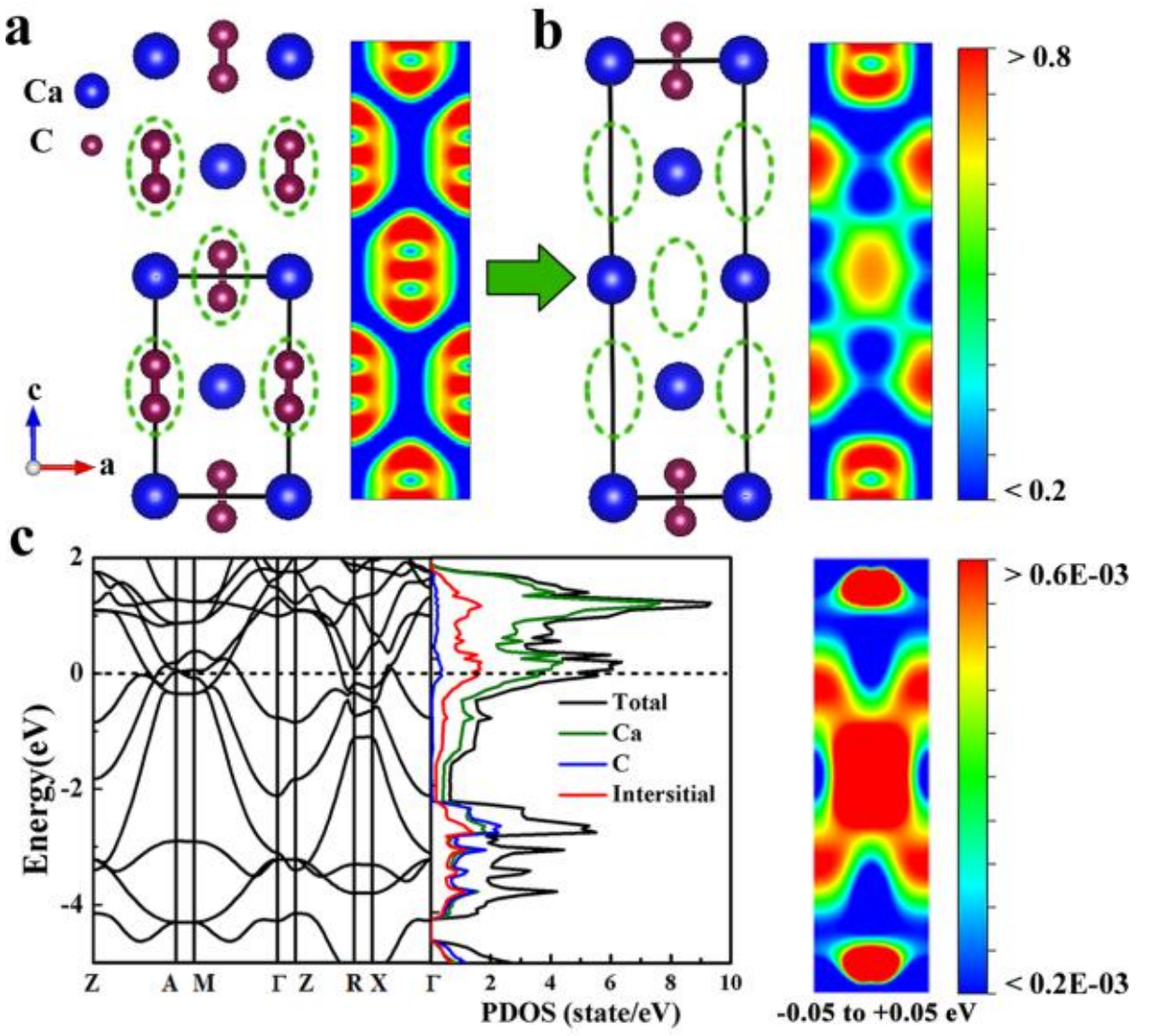

**Figure 4 | Structural and electronic properties of $Ca_2C$ and $CaC_2$.** (a) Crystal structure (left panel) and ELF (right panel) on (0$\bar{1}$2) plane of $CaC_2$. (b) Crystal structure (left panel) and ELF (right panel) on (012) plane of $Ca_2C$. Interstitial regions marked by dashed circles are where excess electrons localize. (c) Electronic properties of $Ca_2C$: band structure and PDOS (left panel), and partial electron density for region near $E_F$ ($|E|<0.05$ eV) on the (012) plane (right panel).

**0D Electrides.** Electron localization of 0D electrides (shown as diamonds in Fig. 3b) is geometrically similar to that of 3D electrides. However, excess electrons in 0D electrides are entirely localized and do not contribute to the conductivity. As a result, these 0D electrides show typically semiconducting/insulating behaviors, which are similar to some of known high-pressure electrides, such as hP4 sodium at 320 GPa[15] and semiconducting Aba2-40 lithium at 70 GPa[16].

$Be_2N$ adopts an *R*3*m* rhombohedral structure, a new electride structure that is also out of the known structure database[32,33]. There are two kinds of N atoms in $Be_2N$: (i) in the first kind, N atoms are six-fold coordinated with Be atoms forming faces shared octahedrons along x-y plane; (ii) N atoms in the second kind are five-fold coordinated with Be atoms forming N centralized hexahedrons (left panel in Fig. 5a). Each hexahedron links to its neighboring hexahedrons or octahedrons by sharing one Be atom. ELF results (Fig. 5a) showed that excess electrons are entirely localized at empty crystallographic positions (yellow spheres in Fig. 5a) in the interstitials. The localized electrons in areas are discrete, which leads to the semi-conductivity of the system (as shown in its band structure in Fig. 5c).

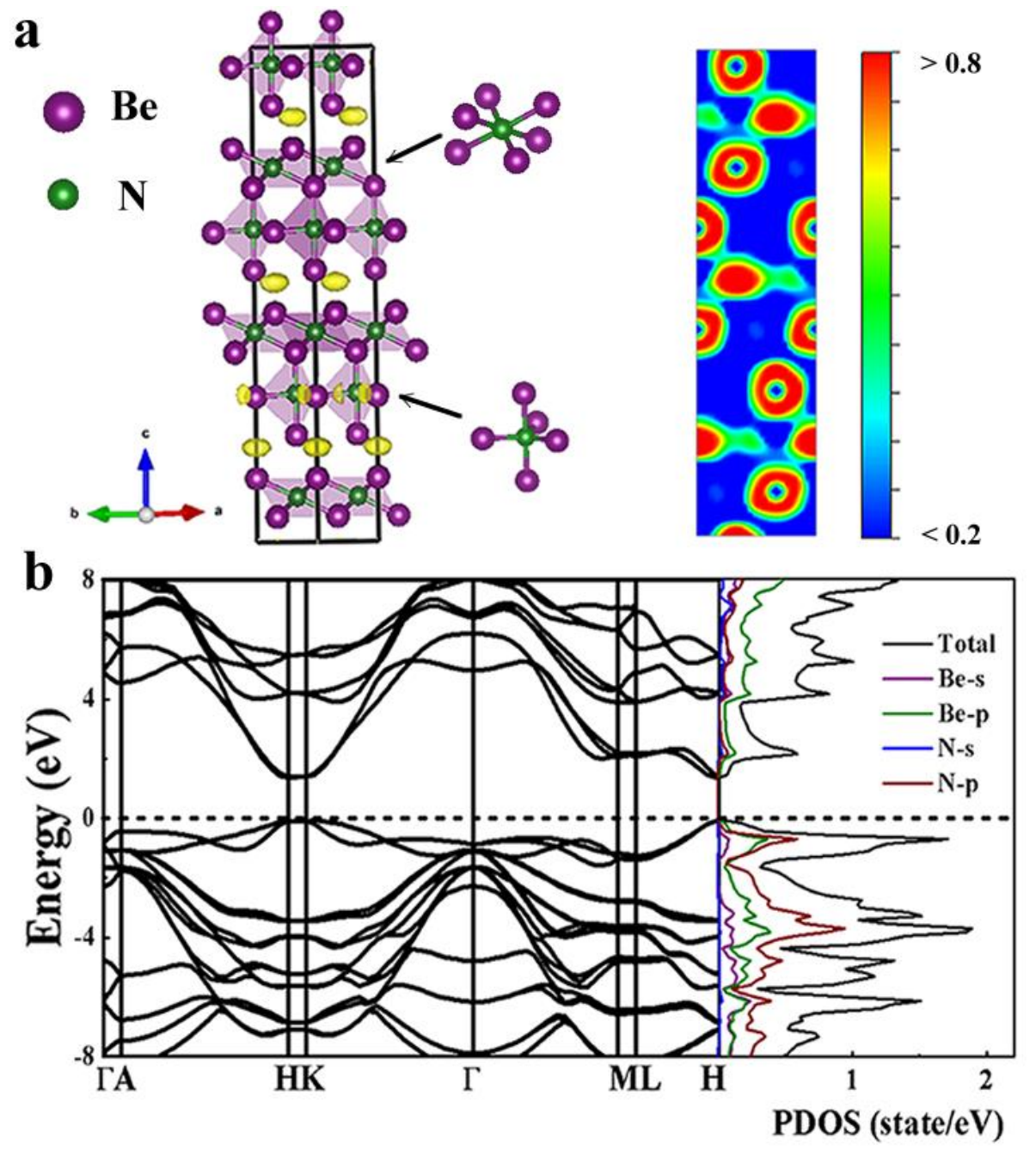

**Figure 5 | Semiconducting $Be_2N$ as a 0D electride.** (a) 3D ELF of $Be_2N$ (left panel) with an isosurface value of 0.83. Six-fold and five-fold coordinated Be atoms are pointed out as inset structural units in the left panel. 2D ELF on (110) plane is plotted out in right panel to show the isolated interstitial localization regions. (b) Band structure and partial density of states (PDOS) of $Be_2N$. The use of HSE06 functional[35,36] corrects the DFT bandgap of 1.51 eV to 1.96 eV. Dashed line indicates the Fermi energy ($E_F$).

**2D Electrides**. In 2D electrides (shown as crosses in Fig. 3b), the excess electrons are confined within interspaces between cationic layers and contribute to the anisotropic conductivity of the system. Note that localized electrons in 2D electrides are not evenly distributed and there are non-nuclear electron maxima at crystallographic positions. 2D electrides have commonly layered structures, however, not all of layered electrides could fall into this category. For example, though $Sc_2N$ and $Y_2N$ share the same layered structure with that of $Ca_2N$, they are not regarded as 2D electrides since the excess electrons are localized so well at the interstitial crystallographic sites and do not contribute to the conductivity.

LaCl[37] adopts a layered *R*-3*m* structure where atomic packing is arranged in a sequence of ABCBA (A, B and C are $Cl^-$ sheets, $La^{3+}$ sheets, and anionic electron layers, respectively) (Fig. 6a). Both La and Cl atoms form a graphene-like arrangement in their respective sheets.

ELF results (top panel in Fig. 6b) showed that the excess electrons are confined in the interlayer interstitial spacings between two La sheets, highlighted in a white dashed circle. Three centers of the maximally localized regions are arrayed into one group that connects each other throughout the layer by delocalized electrons. It is interesting to note that the conducting behaviors of the three centered excess electrons within one group are different. Partial electron density for region near Fermi level indicates that the excess electrons at both ends of one group (circled by a dashed white line in bottom panel in Fig. 6b) are partially localized and contribute to the conductivity of the system, however the excess electrons at the center of the group are well localized without showing any conducting behavior. The dispersive band near Fermi level highlighted in red in band structure (left panel in Fig. 6c) is mainly occupied by interstitial electrons as illustrated by the density of state (right panel in Fig. 6c). The high symmetric lines of Γ-M-K-Γ and A-L-H-A in the reciprocal space depicted in Fig. 6d are indicative of 2D planes of the structure in real space where electrons are localized. The red bands cross Fermi level only along these symmetric lines, showing a clear 2D anisotropic conducting behavior of the system.

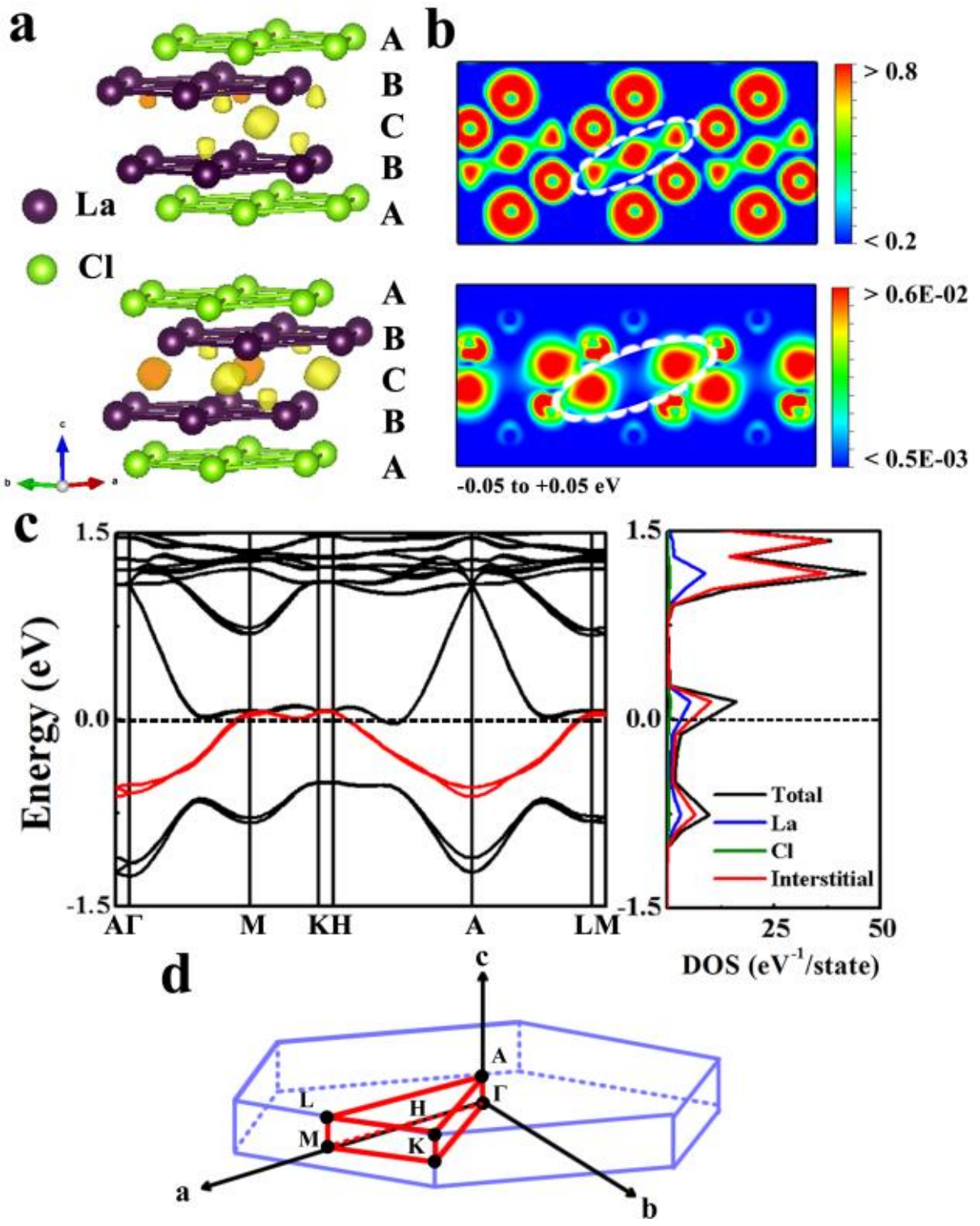

**Figure 6 | Structural and electronic properties of LaCl.** (a) 3D ELF of LaCl with an isosurface value of 0.7. (b) 2D ELF (top panel) on ($\bar{1}\bar{1}0$) plane and partial electron density for region near $E_F$ ($|E|<0.05$ eV) (bottom panel) on the ($\bar{1}\bar{1}0$) plane. (c) Band structure and electron density of state of LaCl. Plotting details are the same as in Figure 5. (d) The high symmetry lines (in red color) in the first Brillouin zone.

## Conclusions

Our developed inverse-design method benefits from the swarm-intelligence based structure search that is able to generate hitherto unknown structure types, and therefore the method is in sharp contrasted to the substitution method relying on the known electride structure of anti-$CdCl_2$ one. We are able to reveal a rich variety of prototype structures of inorganic electrides. 16 structural types out of 89 newly designed electrides are even found to go beyond the known structure databases [32,33] (detailed information for new prototype structures is listed in Supplementary Table S4).

We have extended electrides into more general forms (such as $[A_xB_y]^{n+}$ $(ne)^{n-}$ for binary systems) by considering the formal charges of constitute elements. The two choices of stoichiometry $A_2B$ and AB are just two representative cases. In reality, there exist many alternative stoichiometries as candidates for electrides not only in other binary compounds, but also in ternary, quaternary, and even more complex compounds that are ready for exploration. We believe the findings of electrides will become a general activity especially with the use of our developed inverse-design method. This is in sharp contrast to the current situation of the rarely known inorganic electrides and the greatly limited practical application of inorganic electrides.

Electron-rich ionic systems are chosen in our simulations. This doesn't exclude the possibility on the findings of electrides whose stoichiometry follow the principle of electroneutrality, which is out of scope in this study. We found that electrides can even be extended into electroneutral systems once specific chemical bondings are formed. For example, $Ca_2C$ has been placed into an electride system when C atoms form pairs.

In view of the observed superconductivity in the heavily electron-doped mayenite[38,39], we expect some of our predicted conducting electrides might also show the similar superconducting property. Through explicit calculations of electron-phonon interaction, we found that 3D electrides have potential to be superconducting though with low estimated $T_c$ (e.g., 5 K for $Ca_2Bi$ and 4.17 K for $Sc_2As$). The localized excess electrons can play an important role in the superconductivity of these compounds, which need a further study on the mechanism.

## Methods

**Structure prediction**. Our structure searching simulations are performed through the swarm-intelligence based CALYPSO method[23,24] enabling a global minimization of energy surfaces merging *ab initio* total-energy calculations as implemented in the CALYPSO code. The method has been benchmarked on various known systems. Here, the degree of interstitial electron localization was introduced as the fitness function in the search of electride materials.

**First-principles calculations.** All the electronic structure calculations were performed using density functional theory within the Perdew-Burke-Ernzerhof of generalized gradient approximation as implemented in the Vienna Ab initio Simulation Package (VASP)[40]. The projector-augmented wave (PAW)[41] method was adopted with the PAW potentials taken from the VASP library. The HSE06 hybrid functional[35,36] reproduces well the band gap of semiconductors, thus the HSE06 hybrid functional was applied to revise the band gaps[42] of those semiconducting phases of zero-D electrides calculated by the PBE functional. The plane-wave kinetic energy cutoffs of 800 eV and Monkhorst-Pack Brillouin zone sampling grid with the resolution of $2\pi \times 0.03$ Å$^{-1}$ were chosen to ensure that all the enthalpy calculations are well converged to better than 1 meV/atom. Phonon dispersion and electron-phonon coupling calculations were performed with density functional perturbation theory using the Quantum-ESPRESSO[43] package with kinetic energy cutoffs of 80 Ry. $2\times2\times2$ and $2\times2\times1$ *q*-meshes in the first Brillouin zones were used in the electron-phonon coupling calculations for $Ca_2Sb$ and $Sc_2As$, respectively.

**Author Contributions:**

Y. M. designed the research. Y. Z. performed most of the calculations. Y. Z., H. W., and Y. W. coded the inverse design method into CALYPSO. Y. Z. and Y. M. analyzed the results and wrote the manuscript. All authors commented on the manuscript.

**Acknowledgements**

The authors acknowledge support from National Natural Science Foundation of China under Grants No.11274136 and No.11534003, 2012 Changjiang Scholar of Ministry of Education. L.Z. acknowledges funding support from the Recruitment Program of Global Experts (the Thousand Young Talents Plan).

**Competing financial interests**

The authors declare no competing financial interests.

# **Supplementary Information** for the paper entitled

## "*Inverse Design of Inorganic Electrides*"

Yunwei Zhang[1], Hui Wang[1], Yanchao Wang [1], Lijun Zhang[1], and Yanming Ma[1,*]

[1]*State Key Lab of Superhard Materials, Jilin University, Changchun 130012, China*

**The detailed description on how to get the region $V_{inter}$ where interstitial electron localize well.**

**(a)**

| | | | | | | | | | | | | | | | | | |
|---|---|---|---|---|---|---|---|---|---|---|---|---|---|---|---|---|---|
| 1 H 0.8 | | | | | | | | | | | | | | | | | 2 He |
| 3 Li 1.0 | 4 Be 1.0 | | | | | | | | | | | 5 B 1.8 | 6 C 1.5 | 7 N 1.5 | 8 O 1.5 | 9 F 1.3 | 10 Ne |
| 11 Na 1.0 | 12 Mg 1.0 | | | | | | | | | | | 13 Al 1.0 | 14 Si 2.0 | 15 P 1.9 | 16 S 1.6 | 17 Cl 1.8 | 18 Ar |
| 19 K 1.0 | 20 Ca 1.0 | 21 Sc 1.0 | 22 Ti 1.0 | 23 V 1.0 | 24 Cr 1.0 | 25 Mn 1.0 | 26 Fe 1.0 | 27 Co 1.0 | 28 Ni 1.0 | 29 Cu 1.0 | 30 Zn 1.0 | 31 Ga 1.5 | 32 Ge 2.0 | 33 As 1.9 | 34 Se 1.9 | 35 Br 1.8 | 36 Kr |
| 37 Rb 1.5 | 38 Sr 1.6 | 39 Y 1.0 | 40 Zr 1.0 | 41 Nb 1.0 | 42 Mo 1.0 | 43 Tc 1.0 | 44 Ru 1.0 | 45 Rh 1.0 | 46 Pd 1.0 | 47 Ag 1.0 | 48 Cd 1.0 | 49 In 1.9 | 50 Sn 2.0 | 51 Sb 1.9 | 52 Te 1.9 | 53 I 1.8 | 54 Xe |
| 55 Cs 1.6 | 56 Ba 1.6 | La 1.0 | 72 Hf 1.0 | 73 Ta 1.0 | 74 W 1.0 | 75 Re 1.0 | 76 Os 1.0 | 77 Ir 1.0 | 78 Pt 1.0 | 79 Au 1.0 | 80 Hg 1.0 | 81 Tl 1.9 | 82 Pb 2.0 | 83 Bi 1.9 | 84 Po | 85 At | 86 Rn |
| 87 Fr | 88 Ra | Ac 1.5 | | | | | | | | | | | | | | | |

**(b)**

| | | | | |
|---|---|---|---|---|
| B-B 1.9 | C-C 1.8 | N-N 1.7 | O-O 1.7 | F-F 1.6 |
| | Si-Si 2.5 | P-P 2.4 | S-S 2.3 | Cl-Cl 2.2 |
| | Ge-Ge 2.6 | As-As 2.6 | Se-Se 2.6 | Br-Br 2.5 |
| | | Sb-Sb 2.8 | Te-Te 2.8 | I-I 2.9 |
| | | Bi-Bi 2.8 | | |

**Table S1. Period Tables of Radii (a) and bonds (b).** The interstitial region (***V***) is determined by excluding the ionic spheres and the covalent bonding areas between two atoms. The radii of the ionic sphere $R_{cut}$ (Å) at ambient condition are given in (a). Here the $R_{cut}$ may be larger than Pauling ionic radii in order to repel any anionic region near atoms and have been benchmarked from ELF maps. The covalent bonding lengths (Å) we used in our work are listed in (b). The values listed in two tables are empirical values as reference, user can reset the values in the input file of CALYPSO code. $\boldsymbol{V_{inter}}$ can be got from interstitial region (***V***) and is equal to the volume of interstitial region where electron localize well, that is, where the ELF value larger than 0.75. Here we select 0.75 to distinguish localized region from delocalized region in interstitial region (***V***) for all systems. However, one can change the criterion depending on different systems by themselves.

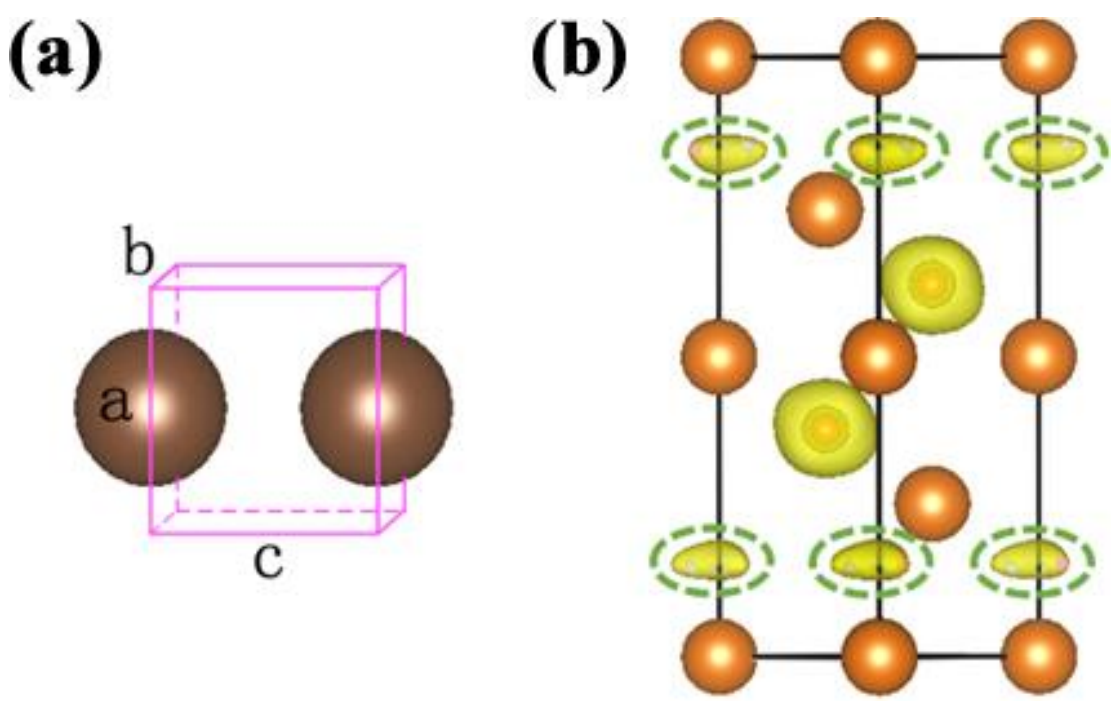

**Figure S1. (a) Covalent bonding areas.** The covalent bonding area of one pair of bonded atoms is a tetragonal lattice depicted in pink lines, with the lattice parameters of a = b = $R_{cut}$ (refer to Table S1. (a)) and c equal to the covalent bond length (refer to Table S1. (b)) between the bonded atoms in a crystal.

**(b) Illustration of the $V_{inter}$ in a crystal.** The interstitial regions where electron localized well ($V_{inter}$) are where the ELF value larger than 0.75 in ELF map. As an illustration, $V_{inter}$ are yellow parts circled by green dashed lines in (b).

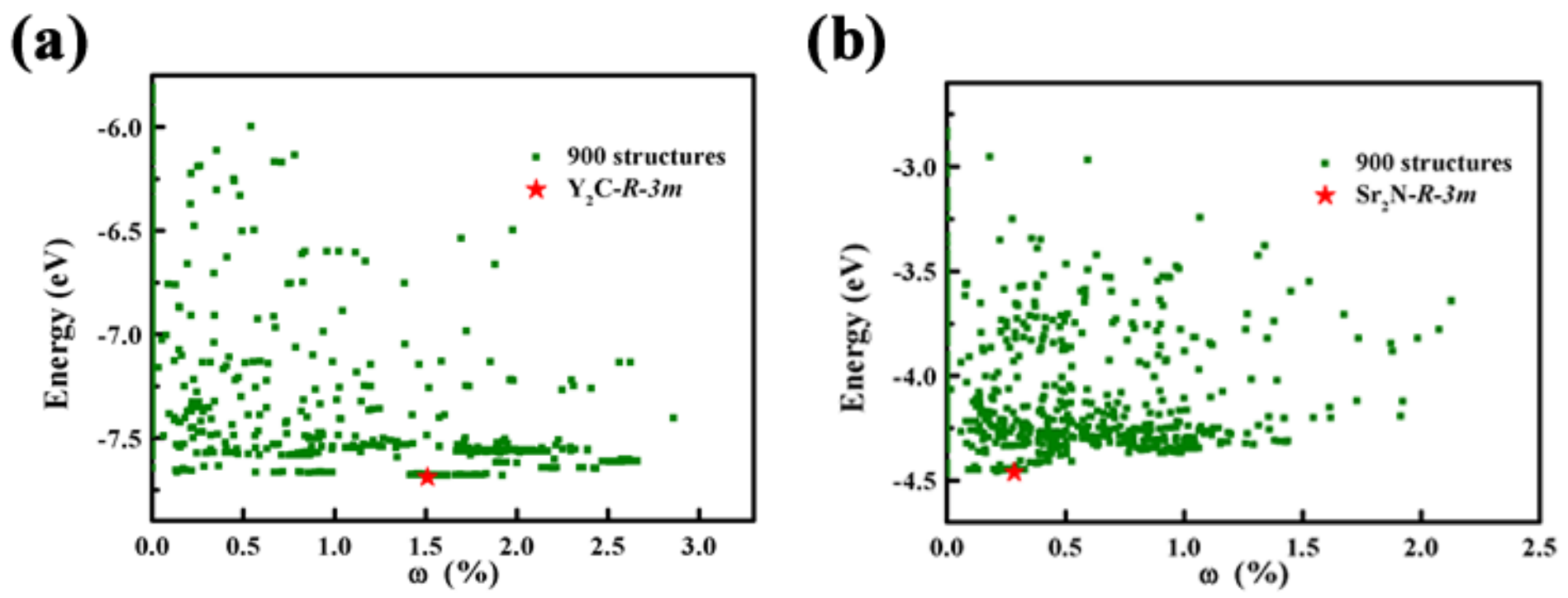

**Figure S2. ω (%) vs. energy maps of $Sr_2N$ (a) and $Y_2C$ (b), respectively.** Green squares are 900 structures produced by CALYPSO code with 30 generations for each system. The desirable *R*3*m* electride structures for $Sr_2N$ and $Y_2C$ are shown as red stars.

| Fomula | Z | Space Group | Prototype structure | a (Å) | b (Å) | c (Å) | ω | $E_{diff}$ | $C_{ref}$ |
|---|---|---|---|---|---|---|---|---|---|
| $Be_2C$ | | | | | | | | | |
| $Be_2N$ | 6 | *R*3*m* | $Be_2N$ | 2.8065 | 2.8065 | 19.8461 | 3.27E-02 | 0.1885 | $Be_3N_2$, Be(s) |
| $Be_2As$ | 6 | *R*3*m* | $Be_2As$ | 4.2070 | 4.2070 | 16.5828 | 9.42E-03 | -0.0611 | As(s), Be(s) |
| $Be_2Sb$ | 6 | *P*3*m*1 | $Be_2Sb$ | 4.2939 | 4.2939 | 7.4351 | 1.29E-02 | 0.0768 | Sb(s), Be(s) |
| $Be_2Bi$ | 6 | *R*-3*m* | $Be_2Bi$ | 3.4688 | 3.4688 | 30.2381 | 1.60E-02 | 0.0174 | Bi(s), Be(s) |
| $Be_2O$ | 6 | *R*3*m* | $Be_2N$ | 2.6114 | 2.6114 | 24.7554 | 5.66E-02 | 0.269 | BeO, Be(s) |
| $Be_2S$ | | | | | | | | | |
| BeCl | 4 | *R*-3*m* | ZrCl | 3.1396 | 3.1396 | 25.7290 | 3.61E-02 | 0.508 | BeCl, Be(s) |
| BeBr | | | | | | | | | |
| $Mg_2C$ | | | | | | | | | |
| $Mg_2N$ | 6 | *Cmcm* | $Mg_2N$ | 3.2606 | 14.2928 | 4.0728 | 7.14E-02 | 0.105 | $Mg_3N_2$, Mg(s) |
| $Mg_2As$ | 6 | *I*4/*mmm* | $La_2Sb$ | 4.1334 | 4.1334 | 14.0811 | 1.25E-02 | 0.077 | $Mg_3As_2$, Mg(s) |
| $Mg_2Sb$ | 3 | *Cmmm* | $ThH_2$ | 12.9459 | 3.4534 | 3.4606 | 5.65E-02 | 0.1313 | $Mg_3Sb_2$, Mg(s) |
| $Mg_2Bi$ | 6 | *Immm* | $Te_2U$ | 13.3562 | 6.3702 | 3.6953 | 1.24E-02 | 0.060175 | $Mg_3Bi_2$, Mg(s) |
| $Mg_2O$ | 6 | *P*-3*m*1 | $Mg_2O$ | 3.0736 | 3.0736 | 10.0092 | 1.37E-02 | 0.1452 | MgO, Mg(s) |
| $Mg_2S$ | 3 | *P*-3*m*1 | $CdI_2$ | 3.6183 | 3.6183 | 3.6183 | 8.28E-02 | 0.138 | MgS, Mg(s) |
| MgCl | 4 | *R*-3*m* | ZrCl | 3.6387 | 3.6387 | 27.4723 | 5.37E-02 | 0.149 | $MgCl_2$, Mg(s) |
| MgBr | 4 | *R*-3*m* | ZrCl | 3.7788 | 3.7788 | 29.4562 | 5.66E-02 | 0.148 | $MgBr_2$, Mg(s) |
| $Ca_2C$ | 6 | *P*4/*mmm* | $Ca_2C$ | 3.7592 | 3.7592 | 12.1695 | 2.63E-02 | -0.00223 | $CaC_2$, Mg(s) |
| $Ca_2N$* | 3 | *R*-3*m* | $CdCl_2$ | 3.6093 | 3.6093 | 19.2631 | 2.78E-03 | 0.09942 | $Ca_3N_2$, Ca(s) |
| $Ca_2As$* | 6 | *I*4/*mmm* | $La_2Sb$ | 4.5992 | 4.5992 | 15.9292 | 4.05E-02 | 0.0436 | $Ca_5As_3$, Ca(s) |
| $Ca_2Sb$* | 6 | *I4/mmm* | $La_2Sb$ | 4.8060 | 4.8060 | 16.8108 | 0.1525753 | -0.0174 | $Ca_5Sb_3$, Ca(s) |
| $Ca_2Bi$* | 6 | *I4/mmm* | $La_2Sb$ | 4.8902 | 4.8902 | 17.0871 | 3.68E-02 | -0.0194 | $Ca_5Bi_3$, Ca(s) |
| $Ca_2O$ | 6 | *R*-3*m* | $Be_2N$ | 3.5669 | 3.5669 | 36.4600 | 4.59E-02 | 0.113 | CaO, Ca(s) |
| $Ca_2S$ | 6 | *R*-3*m* | $Be_2N$ | 4.0084 | 4.0084 | 39.3227 | 8.70E-02 | 0.056708 | CaS, Ca(s) |
| CaCl | 4 | *R*-3*m* | ZrCl | 4.0676 | 4.0676 | 31.1211 | 7.58E-02 | 0.09714 | $CaCl_2$, Ca(s) |
| CaBr | 4 | *R*-3*m* | ZrCl | 4.2287 | 4.2287 | 31.4769 | 6.22E-02 | 0.111 | $CaBr_2$, Ca(s) |
| $Sr_2C$ | 6 | *P4/mmm* | $Ca_2C$ | 4.0603 | 4.0603 | 13.0583 | 6.47E-03 | 0.04215 | $SrC_2$, Sr(s) |
| $Sr_2N$* | 3 | *R*-3*m* | $CdCl_2$ | 3.8617 | 3.8617 | 20.8583 | 1.36E-03 | -0.291 | $SrN_2$, Sr(s) |
| $Sr_2As$* | 6 | *I4/mmm* | $La_2Sb$ | 4.8662 | 4.8662 | 16.9452 | 9.90E-03 | 0.0735 | $Sr_5As_3$, Sr(s) |
| $Sr_2Sb$* | 6 | *I4/mmm* | $La_2Sb$ | 5.0579 | 5.0579 | 17.8723 | 1.23E-02 | 0.0219 | $Sr_5Sb_3$, Sr(s) |
| $Sr_2Bi$* | 6 | *I4/mmm* | $La_2Sb$ | 5.1512 | 5.1512 | 18.1609 | 1.22E-02 | 0.0179 | $Sr_5Bi_3$, Sr(s) |
| $Sr_2O$ | 6 | *R*-3*m* | $Be_2N$ | 3.5669 | 3.5669 | 36.4601 | 5.87E-02 | 0.12945 | SrO, Sr(s) |
| $Sr_2S$ | 6 | *P4/nmm* | $HgI_2$ | 4.2936 | 4.2936 | 12.0230 | 3.00E-02 | 0.0688 | SrS, Sr(s) |
| SrCl | 6 | *P4/nmm* | PbO | 4.8790 | 4.8790 | 6.7727 | 2.66E-02 | 0.264 | $SrCl_2$, Sr(s) |
| SrBr | 6 | *P4/nmm* | SrBr | 4.2159 | 4.2159 | 8.3054 | 2.03E-02 | 0.1195 | $SrBr_2$, Sr(s) |
| $Ba_2C$ | 6 | *P4/mmm* | $Ca_2C$ | 4.318 | 4.318 | 13.8603 | 3.16E-03 | 0.0387 | $BaC_2$, Ba(s) |
| $Ba_2N$* | 3 | *R*-3*m* | $CdCl_2$ | 4.0573 | 4.0573 | 22.8424 | 3.17E-03 | 0.00273 | $Ba_3N$, Ba(s) |
| $Ba_2As$ | 6 | *I4/mmm* | $La_2Sb$ | 5.0932 | 5.0932 | 18.0885 | 6.65E-03 | 0.115 | $Ba_5As_3$, Ba(s) |
| $Ba_2Sb$* | 6 | *I4/mmm* | $La_2Sb$ | 5.2865 | 5.2865 | 19.0667 | 1.74E-03 | 0.02277 | $Ba_5Sb_3$, Ba(s) |
| $Ba_2Bi$* | 6 | *I4/mmm* | $La_2Sb$ | 5.3733 | 5.3733 | 19.3419 | 2.11E-03 | 0.01833 | $Ba_5Bi_3$, Ba(s) |
| $Ba_2O$ | 6 | *P4/nmm* | $Cu_2Sb$ | 5.1875 | 5.1875 | 9.4619 | 4.34E-02 | 0.05275 | BaO, Ba(s) |
| $Ba_2S$ | 6 | *P4/nmm* | $HgI_2$ | 4.5648 | 4.5648 | 12.7880 | 7.19E-03 | 0.068 | BaS, Ba(s) |
| BaCl | 4 | *R*-3*m* | CuI | 5.0894 | 5.0894 | 25.4030 | 2.09E-02 | 0.1075 | $BaCl_2$, Ba(s) |
| BaBr | 6 | *R*-3*m* | CuI | 5.2051 | 5.2051 | 27.483 | 6.73E-03 | 0.105 | $BaBr_2$, Ba(s) |
| $Sc_2C$ | 3 | *R*-3*m* | $CdCl_2$ | 3.3228 | 3.3228 | 16.5998 | 3.48E-02 | -0.0799 | $Sc_4C_3$, Sc(s) |
| $Sc_2N$ | 3 | *R*-3*m* | $CdCl_2$ | 3.2182 | 3.2182 | 15.9749 | 4.58E-02 | -0.01 | ScN, Sc(s) |
| $Sc_2As$ | 6 | *P4/nmm* | $Cu_2Sb$ | 3.9785 | 3.9785 | 7.4978 | 1.73E-02 | 0.0045 | $Sc_5As_3$, Sc(s) |
| $Sc_2Sb$* | 6 | *P4/nmm* | $Cu_2Sb$ | 4.2110 | 4.2110 | 7.8140 | 1.21E-02 | -0.02 | $Sc_5Sb_3$, Sc(s) |
| $Sc_2Bi$ | 6 | *P4/nmm* | $Cu_2Sb$ | 4.2949 | 4.2949 | 7.9737 | 1.29E-02 | -0.103 | ScBi, Sc(s) |
| $Sc_2O$ | 6 | *I*$4_1$ | $TiN_2$ | 4.3774 | 4.3774 | 9.6834 | 7.91E-03 | -0.0482 | $Sc_2O_3$, Sc(s) |
| $Sc_2S$ | 6 | *P4/nmm* | $HgI_2$ | 3.6395 | 3.6395 | 9.2827 | 1.19E-02 | 0.02004 | ScS, Sc(s) |
| ScCl | 4 | *R*-3*m* | ZrCl | 3.5188 | 3.5188 | 30.2238 | 2.31E-02 | 0.0084 | $ScCl_3$, Sc(s) |
| ScBr | 4 | *P*-3*m*1 | ScBr | 3.6156 | 3.6156 | 10.4459 | 1.74E-02 | 0.0454 | $ScBr_3$, Sc(s) |
| $Y_2C$* | 3 | *R*-3*m* | $CdCl_2$ | 3.6117 | 3.6116 | 18.4237 | 1.85E-02 | -0.0961 | $Y_4C_5$, Y(s) |
| $Y_2N$ | 3 | *R*-3*m* | $CdCl_2$ | 3.5219 | 3.5219 | 17.6952 | 2.73E-02 | 0.0503 | YN, Y(s) |
| $Y_2As$ | 6 | *Immm* | $Y_2As$ | 13.0474 | 3.9671 | 6.0464 | 8.19E-03 | 0.0948 | YAs, Y(s) |
| $Y_2Sb$ | 6 | *P4/nmm* | $Cu_2Sb$ | 4.4971 | 4.4971 | 8.4669 | 7.95E-03 | 0.0223 | $Y_5Sb_3$, Y(s) |
| $Y_2Bi$ | 6 | *P4/nmm* | $Cu_2Sb$ | 4.5754 | 4.5754 | 8.6188 | 8.40E-03 | 0.0113 | $Y_5Bi_3$, Y(s) |
| $Y_2O$ | 6 | *P*$4_2$*/nmc* | $HgI_2$ | 3.7612 | 3.7612 | 10.6135 | 4.14E-03 | 0.0825 | $Y_2O_3$, Y(s) |
| $Y_2S$ | 6 | *P4/nmm* | $Y_2S$ | 3.9125 | 3.9125 | 10.1780 | 2.72E-03 | 0.0109 | YS, Y(s) |
| YCl* | 4 | *R*-3*m* | ZrCl | 3.7308 | 3.7308 | 46.1839 | 9.36E-03 | 0.071 | $YCl_3$, Y(s) |
| YBr* | 4 | *P*-3*m*1 | ScBr | 3.8080 | 3.8080 | 14.6172 | 5.67E-03 | 0.0626 | $Y_2Br_3$, Y(s) |
| $La_2C$ | 3 | *P*-3*m*1 | $La_2C$ | 3.7269 | 3.7269 | 6.9088 | 7.63E-04 | -0.03283 | $La_2C_3$, La(s) |
| $La_2N$ | 3 | *R*-3*m* | $CdCl_2$ | 3.6793 | 3.6793 | 20.3756 | 5.08E-03 | -0.00317 | LaN, La(s) |
| $La_2As$ | 6 | *P*6₃/*mmc* | $MoS_2$ | 4.2359 | 4.2359 | 13.5345 | 1.48E-02 | 0.305 | $La_4As_3$, La(s) |
| $La_2Sb$ | 6 | *P4/nmm* | $Cu_2Sb$ | 4.6811 | 4.6811 | 9.1226 | 1.89E-03 | -0.019 | LaSb, La(s) |
| $La_2Bi$ | 3 | *R*-3*m* | $CdCl_2$ | 4.4640 | 4.4640 | 21.1735 | 2.16E-03 | 0.281 | $La_5Bi_3$, La(s) |
| $La_2O$ | 3 | *R*-3*m* | $CdI_2$ | 3.7951 | 3.7951 | 6.2679 | 6.29E-03 | 0.1811 | LaO, La(s) |
| $La_2S$ | 6 | *R*3*m* | $La_2S$ | 3.9462 | 3.9462 | 40.6260 | 7.92E-02 | 0.0755 | LaS, La(s) |
| LaCl* | 4 | *R*-3*m* | ZrCl | 4.0654 | 4.0654 | 30.3248 | 9.73E-03 | 0.055 | $LaCl_3$, La(s) |
| LaBr* | 4 | *R*-3*m* | ZrCl | 4.1333 | 4.1333 | 29.7150 | 9.01E-03 | 0.1015 | $LaBr_2$, La(s) |
| $Zr_2C$ | 3 | *R*-3*m* | $CdCl_2$ | 3.3478 | 3.3478 | 16.1704 | 9.51E-03 | -0.057 | ZrC, Zr(s) |
| $Zr_2N$ | 3 | *R*-3*m1* | $CdI_2$ | 3.2919 | 3.2919 | 5.2698 | 6.86E-03 | -0.0504 | ZrN, Zr(s) |
| $Zr_2As$ | 3 | *I4/mmm* | $MoSi_2$ | 3.5674 | 3.5674 | 9.5248 | 2.11E-02 | 0.354 | $Zr_3As_2$, Zr(s) |
| $Zr_2Sb$ | 4 | *I4/mmm* | $MoSi_2$ | 3.7492 | 3.7492 | 9.6087 | 1.46E-02 | 0.0481 | $Zr_5Sb_3$, Zr(s) |
| $Zr_2Bi$ | 4 | *I4/mmm* | $MoSi_2$ | 3.8123 | 3.8123 | 9.7705 | 1.43E-02 | -0.0377 | ZrBi, Zr(s) |
| $Zr_2O$ | 6 | *Pnnm* | $FeS_2$ | 5.1766 | 5.1766 | 3.2179 | 1.30E-02 | -0.3018 | ZrO, Zr(s) |
| $Zr_2S$* | 3 | *P*-3*m*1 | $CdI_2$ | 3.5384 | 3.5384 | 5.6365 | 2.00E-02 | -0.175 | ZrS, Zr(s) |
| ZrCl* | 4 | *R*-3*m* | ZrCl | 3.4272 | 3.4272 | 30.3364 | 2.26E-02 | -0.133 | $ZrCl_3$, Zr(s) |
| ZrBr* | 4 | *R*-3*m* | ZrCl | 3.5258 | 3.5258 | 31.5407 | 1.96E-02 | -0.0757 | $ZrBr_3$, Zr(s) |
| $Hf_2C$ | 3 | *R*-3*m* | $CdCl_2$ | 3.2869 | 3.2869 | 15.7881 | 4.55E-03 | -0.094 | HfC, Hf(s) |
| $Hf_2N$ | 3 | *P*-3*m*1 | $CdI_2$ | 3.2333 | 3.2333 | 5.1616 | 1.60E-02 | -0.1358 | HfN, Hf(s) |
| $Hf_2As$ | 6 | *Immm* | $Te_2U$ | 3.4908 | 5.6141 | 12.1631 | 1.00E-02 | 0.0281 | $Hf_3As_2$, Hf(s) |
| $Hf_2Sb$ | 3 | *I4/mmm* | $MoSi_2$ | 3.6988 | 3.6988 | 9.5574 | 1.75E-02 | -0.00421 | $Hf_5Sb_3$, Hf(s) |
| $Hf_2Bi$ | 3 | *I4/mmm* | $MoSi_2$ | 3.7547 | 3.7547 | 9.7950 | 1.64E-02 | -0.06017 | HfBi, Hf(s) |
| $Hf_2O$ | 3 | *P*-3*m*1 | $CdI_2$ | 3.2187 | 3.2187 | 5.1306 | 4.28E-02 | -0.13738 | $HfO_2$, Hf(s) |
| $Hf_2S$* | 6 | *P*6₃/*mmc* | $NbS_2$ | 3.3748 | 3.3748 | 11.7720 | 3.20E-02 | -0.3333 | $HfS_2$, Hf(s) |
| HfCl* | 4 | *R*-3*m* | ZrCl | 3.3704 | 3.3704 | 29.6932 | 2.50E-02 | -0.1578 | $HfCl_4$, Hf(s) |
| HfBr* | 4 | *R*-3*m* | ZrCl | 3.5067 | 3.5067 | 31.6268 | 2.26E-02 | -0.112 | $HfBr_4$, Hf(s) |
| $Al_2C$ | 3 | *P*-3*m*1 | $CdI_2$ | 3.1047 | 3.1047 | 4.4460 | 3.64E-02 | 0.0709 | $Al_4C_3$, Al(s) |
| $Al_2N$ | 6 | *I*-4*m*2 | $Al_2N$ | 3.117 | 3.117 | 17.1591 | 8.71E-02 | 0.0826 | AlN, Al(s) |
| $Al_2As$ | 6 | *I*-4*m*2 | $Al_2N$ | 3.9138 | 3.9138 | 19.3565 | 5.75E-02 | 0.229 | AlAs, Al(s) |
| $Al_2Sb$ | 6 | *I*-4*m*2 | $Al_2N$ | 4.2602 | 4.2602 | 19.6040 | 6.82E-02 | 0.275 | AlSb, Al(s) |
| $Al_2Bi$ | | | | | | | | | |
| $Al_2O$ | 2 | *R*3*m* | $Al_2O$ | 2.8912 | 2.8912 | 10.4176 | 1.94 E-03 | 0.131 | $Al_2O_3$, Al(s) |
| $Al_2S$ | 6 | *P*-3*m*1 | $Al_2S$ | 3.3924 | 3.3924 | 12.9981 | 3.05E-02 | 0.165 | $Al_2S_3$, Al(s) |
| AlCl | 4 | *Pmmn* | AlCl | 3.1735 | 3.5344 | 8.6712 | 3.19E-02 | 0.24 | $AlCl_3$, Al(s) |
| AlBr | | | | | | | | | |

**Table S2. Prototype structure, relaxed lattice data, degree of interstitial electron localization, and formation energy with respect to existing compounds of all designed binary electrides.** a, b, and c are crystal lattice parameters of their unit cell, respectively. ω is the degree of interstitial electron localization. $E_{diff}$ is the formation energy with respect to the existing stable compounds ($C_{ref}$). Compounds with asterisk (*) have already been synthesized.

| Elements | Ratio (%) | Elements | Ratio (%) | Elements | Ratio (%) | Elements | Ratio (%) |
|---|---|---|---|---|---|---|---|
| $Li_4N$ | 83.1 | $Tl_2N$ | 0.0 | $Hf_2N$ | 29.2 | $Co_2N$ | 0.0 |
| $Na_4N$ | 62.2 | $Ge_2N$ | 0.0 | $V_2N$ | 0.6 | $Rh_2N$ | 0.0 |
| $K_4N$ | 23.1 | $Sn_2N$ | 0.0 | $Nb_2N$ | 0.6 | $Ir_2N$ | 0.0 |
| $Rb_4N$ | 15.6 | $Pb_2N$ | 0.0 | $Ta_2N$ | 15.6 | $Ni_2N$ | 0.0 |
| $Cs_4N$ | 2.3 | $Sb_2N$ | 0.0 | $Cr_2N$ | 3.3 | $Pd_2N$ | 0.0 |
| $Be_2N$ | 74.1 | $Bi_2N$ | 0.0 | $Mo_2N$ | 0.2 | $Pt_2N$ | 0.0 |
| $Mg_2N$ | 51.8 | $Po_2N$ | 0.0 | $W_2N$ | 0,2 | $Cu_2N$ | 0.0 |
| $Ca_2N$ | 87.7 | $Sc_2N$ | 87.6 | $Mn_2N$ | 0.0 | $Ag_2N$ | 0.0 |
| $Sr_2N$ | 65.1 | $Y_2N$ | 71.5 | $Tc_2N$ | 0.0 | $Au_2N$ | 0.0 |
| $Ba_2N$ | 37.7 | $La_2N$ | 20.2 | $Re_2N$ | 0.0 | $Zn_2N$ | 0.0 |
| $Al_2N$ | 87.5 | $Ac_2N$ | 1.8 | $Fe_2N$ | 1.3 | $Cd_2N$ | 0.0 |
| $Ga_2N$ | 0.2 | $Ti_2N$ | 3.5 | $Ru_2N$ | 0.0 | $Hg_2N$ | 0.0 |
| $In_2N$ | 0.0 | $Zr_2N$ | 13.8 | $Os_2N$ | 0.0 | | |

**Table S3. Chemical stoichiometries of nitrides in the model system of metal nitrides and the ratios of electride structures over all structures generated by CALYPSO code (formation probability of electrides).**

| Prototype | Space group | Z | Wyckoff position |
|---|---|---|---|
| $Be_2N$ | *R*3*m* | 6 | Be1: 3a (0, 0, 5885), Be4: 3a (0, 0, 0.1679)<br>Be7: 3a (0, 0, 4234), Be10: 3a (0, 0, 0)<br>N13: 3a (0, 0, 0.2931), N16: 3a (0, 0, 0.0844) |
| $Be_2As$ | *R*3*m* | 6 | Be1: 3a (0, 0, -0.4342), Be2: 9b (0.1648, -0.1648, -0.3145)<br>As5: 3a (0, 0, -0.0967), As6: 3a (0, 0, -0.5794) |
| $Be_2Sb$ | *P*3*m*1 | 2 | Be1: 3d (0.4983, -0.0035, 0.1496), Be2: 1b (0.3333, 0.6667, 0.8994)<br>Sb5: 1c(0.6667, 0.3333, 0.8493), Sb6: 1a(0, 0, 0.3388) |
| $Mg_2N$ | *Cmcm* | 4 | Mg1: 4c(1, -0.0857, 0.25), Mg3: 4c(0.5, 0.2165, 0.25)<br>N5: 4c(0.5, 0.1815, 0.75) |
| $Mg_2O$ | *P*-3*m*1 | 2 | Mg1: 2d(0.3333. 0.6667, 0.2546), Mg2: 1b(0, 0, 0.5)<br>Mg3: 1a(0, 0, 0)<br>O5: 2d(0.6667, 0.3333, 0.3790) |
| $Ca_2C$ | *P4/mmm* | 2 | Ca1: 2f (0, 0, 0.5), Ca2: 2h (0.5, 0.5, 0.7457)<br>Ca3: 2h (0.5, 0.5, 0.2543), Ca4: 1a (1, 1, 1)<br>C1: 2h (0.5, 0.5, 0.0537), C2: 2h (0.5, 0.5, 0.9463) |
| SrBr | *P4/nmm* | 2 | Sr1: 2c(0, 0.5, 0.7068)<br>Br3: 2c(0, 0,5, 0.1370) |
| ScBr | *P*-3*m*1 | 2 | Sc1: 2d(0.3333, 0.6667, 0.8799)<br>Br3: 2c(0, 0, 0.2879) |
| $Y_2As$ | *Immm* | 4 | Y1: 4j(0, 0.5, 0.3007), Y2: 4g(0, 0.2601, 0) |
| | | | As9: 4i(0, 0, 0.6413) |
| $Y_2S$ | *P4/nmm* | 2 | Y1: 2c(0.5, 0, 0.6035), Y3: 2c(0.5, 0, 0.1404)<br>S5: 2c(0.5, 0, 0.8735) |
| $La_2C$ | *P*-3*m*1 | 1 | La1: 2d(-0.3333, -0.6667, 0.2092)<br>C5: 1a(0, 0, 0) |
| $La_2S$ | *R*3*m* | 6 | La1: 3a(0, 0, 0.2686), La2: 3a(0, 0, 0.1162),<br>La3: 3a(0, 0, 0.6923), La4: 3a(0, 0, 0.8594)<br>S5: 3a(0, 0, 0.4032), S6: (0, 0, 0.9813) |
| $Al_2N$ | *I-4m*2 | 4 | Al1: 4f(0.5, 0, 0.6221), Al2: 2a(0.5, 0.5, 0.5),<br>Al5: 2d(0, 0.5, 0.75)<br>N9: 4e(0.5, 0.5, 0.8137) |
| $Al_2O$ | *R*-3*m* | 6 | Al1: 3b(0, 0, 0.5), Al2: (0, 0, 0)<br>Al3: 6c(0, 0, 0.7610)<br>O5: 6c(0, 0, 0.3020) |
| $Al_2S$ | *P*-3*m*1 | 2 | Al1: 2d(0.6667, 0.3333, 0.0767), Al2: (0, 0, 0.7548)<br>S5: 2d(0.6667, 0.3333, 0.6641) |
| AlCl | *Pmmn* | 2 | Al1: 2a(0, 0, 0.5859)<br>Cl3: 2b(0.5, 0, 0.7849) |

**Table S4. Structural information of 16 new prototype structures.**